\documentclass[onecolumn,showpacs,floatfix,12pt,nofootinbib]{revtex4}
\usepackage[dvips]{graphics,color}
\usepackage{epsfig}\usepackage{float}
\RequirePackage{amssymb}
\usepackage{makeidx}
\usepackage[brazil]{babel}
\usepackage[latin1]{inputenc}
\usepackage{graphicx}
\usepackage{indentfirst}
\usepackage{color}
\newcommand{\be}{\begin{equation}}
\newcommand{\ee}{\end{equation}}

\newcommand{\ba}{\begin{eqnarray}}
\newcommand{\ea}{\end{eqnarray}}

\begin{document}

\title{Gauss-Codazzi formalism to brane-world within Brans-Dicke theory}

\author{M. C. B. Abdalla$^{1}$}
\email{mabdalla@ift.unesp.br}
\author{M. E. X. Guimar\~aes$^{2}$}
\email{emilia@if.uff.br}
\author{J. M. Hoff da Silva$^{1}$}
\email{hoff@ift.unesp.br}

\affiliation{1. Instituto de F\'{\i}sica Te\'orica, Universidade
Estadual Paulista, Rua Pamplona 145 01405-900 S\~ao Paulo, SP,
Brazil}

\affiliation{2. Instituto de F\'{\i}sica, Universidade Federal
Fluminense, Niter\'oi-RJ, Brazil}


\begin{abstract}
We apply the Gauss-Codazzi formalism to brane-worlds within the
framework of Brans-Dicke gravity. The compactification is taken
from six to five dimensions in order to formalize brane-world
models with hybrid compactification in scalar-tensor theories.

\end{abstract}
\maketitle
\section{Introduction}
Advances in the establishment of string theory point to a
multidimensional world \cite{HW}. This claim, from among other
problems such as the hierarchy one, originated several works in
which our universe is understood as a membrane (the brane), or a
submanifold, embedded into a higher dimensional spacetime (the
bulk) \cite{muitos}. In the main compactification of M-theory
there is a quite interesting symmetry of the orbifold topology
which is given by the $\mathbb{Z}_{2}$ discrete group. From the
point of view of gravity, this symmetry provides a simple solution
for the extrinsic curvature on the brane and enables the full
construction of the Gauss-Codazzi formalism for a brane-world with
one extra noncompact dimension \cite{JP}. This symmetry is also
useful to analyze chiral fermions on the brane \cite{SUNDRUM}.
However it is no longer a necessity in both problems. In the
brane-world domain, there is a generalization of the Gauss-Codazzi
formalism without a $\mathbb{Z}_{2}$ symmetry \cite{JPII} and, it
seems that index theorems can lead to a good approach for the
chirality problem. In other words, $\mathbb{Z}_{2}$ symmetry seems
to be desirable but not indispensable.

On the other hand, Brans-Dicke gravity \cite{BD} is part (just by
a relabelling of the Brans-Dicke parameter\footnote{In fact, such
duality depends on the specific compactification scenario, see
\cite{Perivo} for instance.} $w$ \cite{novos}) of gravity
recovered from string theory at low energy. In such a framework,
recent advances using cosmic string as topological defects to
generate the compactification \cite{AG} suggest that the resultant
setup can be given in terms of a hybrid compactification scenario,
i. e., when there is a noncompact extra dimension in the bulk and,
at the same time, a compact one on brane. So, the picture is the
compactification from six to five dimensions, where there is a
small compact dimension $S^{1}$ at each point on the brane. This
type of scenario is potentially interesting because it can provide
a good approach to explain the hierarchy problem, due to the
hybrid compactification, and also suggests a candidate to dark
matter. However, this last claim needs further investigation in
warped geometries.

As we mentioned above, the Brans-Dicke theory has a strong
relation with low energy gravitation recovered from string theory.
Besides, the scalar field opens a new possibility in the scale
adjustment of the Higgs mechanism. Apart of this, a rigorous
formulation of braneworld models needs to treat the brane as a
gravitational object with a non-zero tension. It is a requirement
to study gravitational systems, as black holes for instance, on
the brane. Of course, the presence of the new scalar field in the
bulk can also bring new possibilities in such systems. In this
vein, the manipulation and extension of usual Gauss-Codazzi
formalism to scalar tensorial theories is quite necessary. It was
done to braneworlds in usual General Relativity \cite{JP}, but the
additional scalar field turns the manipulation of the equations
more complicated and, as we will see later, introduces important
differences in the final result.

This work is an attempt to establish the Gauss-Codazzi formalism
for this kind of model. As we will see, the Brans-Dicke scalar
field plays an important role in the effective cosmological
constant on the brane, as well as in the Newton's constant. In a
few words, this work is a generalization of Gauss-Codazzi
equations for Brans-Dicke theory from six to five dimensions
inspired by the results found in \cite{AG} and motivated by the
above reasons. The paper is organized as follows: in Section II we
apply the Gauss and the Codazzi equations for the brane-world in
question and relate the Einstein equations on the brane with some
bulk quantities using $\mathbb{Z}_{2}$ symmetry, in a similar
fashion of reference \cite{JP}. In the last section we conclude
calling attention for the next step of this program, which is
working without $\mathbb{Z}_{2}$ symmetry. In the Appendix we show
the matching condition for Brans-Dicke theory.

\section{Gauss-Codazzi in the Brans-Dicke Theory}

We treat the brane as a submanifold, in five dimensions, out of a
six dimensional manifold. It is important to remark that one of
the five brane dimensions is compactified into a $S^{1}$ topology,
so it is a codimension one scenario. The importance of this fact
is that in higher codimension models the Gauss-Codazzi formalism
is no longer useful\footnote{Apart of the regularization mechanism
introduced in \cite{JPII}.}, since we need some minimal regularity
near the brane in order to implement it \cite{RUTH}.

\subsection{Notation and Conventions}

Basically we follow the notation and conventions used in
\cite{JP}. The covariant derivative associated to the bulk is
labelled by $\nabla_{\mu}$ while the one associated to the brane
is $D_{\mu}$. The induced metric on the brane is given by
$q_{\mu\nu}=g_{\mu\nu}-n_{\mu}n_{\nu}$, where $n_{\mu}$ is the
unitary normal vector to the brane. In such terms, the Gauss
equation reads
\be{^{(5)}\!R^{\alpha}_{\beta\gamma\delta}=^{(6)}\!\!\!R^{\mu}_{\nu\rho\sigma}q_{\mu}^{\;\alpha}q_{\beta}^{\;\nu}q_{\gamma}^{\;\rho}q_{\delta}^{\;\sigma}+
K^{\alpha}_{\;\gamma}K_{\beta\delta}-K^{\alpha}_{\;\delta}K_{\beta\gamma}},\label{1}\ee
while the Codazzi equation is
\be{D_{\nu}K_{\mu}^{\;\nu}-D_{\mu}K=^{(6)}\!\!\!R_{\rho\sigma}n^{\sigma}q_{\mu}^{\;\rho},\label{2}
}\ee where
$K_{\mu\nu}=q_{\mu}^{\;\alpha}q_{\nu}^{\;\beta}\nabla_{\alpha}n_{\beta}$
is the extrinsic curvature. From the equation (\ref{1}) it is easy
to find the Einstein's equation in five dimensions in terms of the
bulk quantities. The Ricci tensor is
\be{^{(5)}\!R_{\beta\delta}=^{(6)}\!\!\!R_{\nu\sigma}q_{\beta}^{\;\nu}q_{\delta}^{\;\sigma}-^{(6)}\!\!R^{\mu}_{\nu\rho\sigma}n_{\mu}n^{\rho}q_{\beta}^{\;\nu}q_{\delta}^{\;\sigma}
+KK_{\beta\delta}-K^{\gamma}_{\;\delta}K_{\beta\gamma}},\label{3}\ee
and the scalar of curvature is given by
\be{^{(5)}\!R=^{(6)}\!\!\!R_{\nu\sigma}q^{\nu\sigma}-^{(6)}\!\!R^{\mu}_{\nu\rho\sigma}n_{\mu}n^{\rho}q^{\nu\sigma}+K^{2}-K^{\alpha\beta}K_{\alpha\beta}
}.\label{4}\ee Then, the Einstein's equation is given by \ba
^{(5)}\!G_{\beta\delta}&=&\left.^{(6)}\!G_{\nu\sigma}q_{\beta}^{\;\nu}q_{\delta}^{\;\sigma}+^{(6)}\!\!\!R_{\nu\sigma}n^{\nu}n^{\sigma}q_{\beta\delta}+KK_{\beta\delta}
-K_{\delta}^{\;\gamma}K_{\beta\gamma}\right.\nonumber\\&-&\left.
\frac{1}{2}q_{\beta\delta}(K^{2}-K^{\alpha\gamma}K_{\alpha\gamma})-\tilde{E}_{\beta\delta},\right.
\label{5}\ea where
$\tilde{E}_{\beta\delta}=^{(6)}\!\!R^{\mu}_{\;\nu\rho\sigma}n_{\mu}n^{\rho}q_{\beta}^{\;\nu}q_{\delta}^{\;\sigma}$.

It is well known that the Riemann, Ricci and Weyl tensors are
related among themselves, in an arbitrary dimension $(n)$, by \ba
^{(n)}\!R_{\alpha\beta\mu\nu}&=&\left.
^{(n)}\!C_{\alpha\beta\mu\nu}+\frac{2}{n-2}\Big(\;
 ^{(n)}\!R_{\alpha[\mu}g_{\nu]\beta}-^{(n)}\!\!\!R_{\beta[\mu}g_{\nu]\alpha}
\Big)\right.\nonumber\\&-&\left.\frac{2}{(n-1)(n-2)}\;
^{(n)}\!Rg_{\alpha[\mu}g_{\nu]\beta}.\right. \label{9}\ea After
some manipulation, the term $\tilde{E}_{\beta\delta}$ in (\ref{5})
can be written as
\be{\tilde{E}_{\beta\delta}=E_{\beta\delta}+\frac{1}{2}\Big(\;
 ^{(6)}\!R^{\mu}_{\;\rho}n_{\mu}n^{\rho}q_{\beta\delta}+
^{(6)}\!\!R_{\nu\sigma}q_{\beta}^{\;\nu}q_{\delta}^{\;\sigma}\Big)-\frac{1}{10}\;
^{(6)}\!Rq_{\beta\delta}},\label{pc}\ee where
$E_{\beta\delta}=^{(6)}\!C^{\mu}_{\;\nu\rho\sigma}n_{\mu}n^{\rho}q_{\beta}^{\;\nu}q_{\rho}^{\;\sigma}$.
In terms of this new quantity the equation (\ref{5}) reads \ba
^{(5)}\!G_{\beta\delta}&=&\left.\frac{1}{2}\;
^{(6)}\!G_{\nu\sigma}q_{\beta}^{\;\nu}q_{\delta}^{\;\sigma}-
\frac{1}{10}\;^{(6)}\!Rq_{\beta\delta}-\frac{1}{2}\;^{(6)}\!R_{\nu\sigma}q^{\nu\sigma}q_{\beta\delta}+KK_{\beta\delta}
-K^{\gamma}_{\;\delta}K_{\beta\gamma}\right.\nonumber\\&-&\left.\frac{1}{2}q_{\beta\delta}(K^{2}-K^{\alpha\gamma}K_{\alpha\gamma})-E_{\beta\delta}.\right.\label{prep}\ea

\subsection{Notation and Conventions in Brans-Dicke Theory}

Now, the generalization to the Brans-Dicke gravity means to
express the right-hand side of (\ref{prep}) in terms of the scalar
field of such a theory. The Einstein-Brans-Dicke equation is \ba
^{(6)}\!G_{\mu\nu}&=&\left.\frac{8\pi}{\phi}T_{M\mu\nu}+\frac{w}{\phi^{2}}\Big(\nabla_{\mu}\phi\nabla_{\nu}\phi-
\frac{1}{2}g_{\mu\nu}\nabla_{\alpha}\phi\nabla^{\alpha}\phi\Big)\right.\nonumber
\\&+&\left.\frac{1}{\phi}\Big(\nabla_{\mu}\nabla_{\nu}\phi-g_{\mu\nu}\Box^{2}\phi\Big),\right.\label{6}\ea
where $\phi$ is the Brans-Dicke scalar field, $w$ a dimensionless
parameter and $T_{M\mu\nu}$ the matter energy-momentum tensor.
Everything except $\phi$ and gravity, in the bulk. The scalar
equation of Brans-Dicke theory is given by \ba
\Box^{2}\phi=\frac{8\pi}{3+2w}T_{M}.\label{scalar} \ea

From the equations (\ref{6}) and (\ref{scalar}) one can find the
scalar of curvature of the bulk, as well as, the Ricci tensor. The
scalar has the form
\be{^{(6)}\!R=\frac{-8\pi}{\phi}\Bigg(\frac{w-1}{3+2w}\Bigg)T_{M}+\frac{w}{\phi^{2}}\nabla_{\alpha}\phi\nabla^{\alpha}\phi}\label{7}\ee
and the Ricci tensor
\be{^{(6)}\!R_{\mu\nu}=\frac{8\pi}{\phi}\Bigg[T_{M\mu\nu}-\frac{1}{2}
g_{\mu\nu}\Bigg(\frac{1+w}{3+2w}\Bigg)T_{M}\Bigg]+\frac{1}{\phi}\nabla_{\mu}\nabla_{\nu}\phi
+\frac{w}{\phi^{2}}\nabla_{\mu}\phi\nabla_{\nu}\phi}.\label{8}\ee

Substituting the equations (\ref{6}), (\ref{7}) and (\ref{8}) into
(\ref{prep}) we find the first step to the generalization of the
Gauss-Codazzi formalism to the Brans-Dicke theory, encoded in the
following equation \ba
^{(5)}\!G_{\beta\delta}&=&\left.\frac{1}{2}\Bigg[
\frac{8\pi}{\phi}T_{M\nu\sigma}+\frac{1}{\phi}\nabla_{\nu}\nabla_{\sigma}\phi+\frac{w}{\phi^{2}}\nabla_{\nu}\phi\nabla_{\sigma}\phi
\Bigg](q_{\beta}^{\;\nu}q_{\delta}^{\;\sigma}-q^{\nu\sigma}q_{\beta\delta})\right.\nonumber\\&+&\left.\frac{2\pi}{5\phi}q_{\beta\delta}
T_{M}\Bigg(\frac{13+27w}{3+2w}\Bigg)-\frac{7w}{20\phi^2}q_{\beta\delta}\nabla_{\alpha}\phi\nabla^{\alpha}\phi+KK_{\beta\delta}-K^{\gamma}_{\;\delta}
K_{\beta\gamma}\right.\nonumber\\&-&\left.\frac{1}{2}q_{\beta\delta}(K^{2}-K^{\alpha\gamma}K_{\alpha\gamma})-E_{\beta\delta}.\right.\label{NG}\ea
The Codazzi equation (\ref{2}) together with (\ref{8}) gives
\be{D_{\nu}K_{\mu}^{\;\nu}-D_{\mu}K=\Bigg[\frac{8\pi}{\phi}T_{M\rho\sigma}+\frac{1}{\phi}\nabla_{\rho}\nabla_{\sigma}\phi+\frac{w}{\phi^{2}}
\nabla_{\rho}\phi\nabla_{\sigma}\phi\Bigg]n^{\sigma}q_{\mu}^{\;\rho}}.\label{NC}\ee

The equations (\ref{NG}) and (\ref{NC}) summarizes this stage. To
extract information about the system, we have to compute some
quantity on the brane, or for better saying, taking the limit of
the extra dimensions tending to the brane. The extrinsic curvature
is an important tool for the matching conditions. In order to
guarantee the sequential reading of the paper, we derive these
equations in the Appendix.

The effect of imposing the $\mathbb{Z}_{2}$ symmetry on the brane
(or on each brane, in a multi-brane scenario) is to change the
signal of the $n_{\alpha}$ vector across the brane, and
consequently, to change the signal of the extrinsic curvature.
Then, taking the equation (\ref{ap15}) of the Appendix into
account we have
\be{K_{\mu\nu}^{+}=-K_{\mu\nu}^{-}=\frac{4\pi}{\phi}\Bigg(-T_{\mu\nu}+\frac{q_{\mu\nu}(1+w)T}{2(3+2w)}
\Bigg)},\label{29}\ee and
\be{K^{+}=K^{-}=\frac{2\pi}{\phi}\Bigg(\frac{w-1}{3+2w}\Bigg)T}.\label{30}\ee

Plugging the equations (\ref{29}) and (\ref{30}) into equation
(\ref{NG}) we have, after some algebra, the following result \ba
^{(5)}\!G_{\beta\delta}&=&\left.\frac{1}{2}\Bigg[
\frac{8\pi}{\phi}T_{M\nu\sigma}+\frac{1}{\phi}\nabla_{\nu}\nabla_{\sigma}\phi+\frac{w}{\phi^{2}}\nabla_{\nu}\phi\nabla_{\sigma}\phi
\Bigg](q_{\beta}^{\;\nu}q_{\delta}^{\;\sigma}-q^{\nu\sigma}q_{\beta\delta})\right.\nonumber\\&+&\left.\frac{2\pi}{5\phi}q_{\beta\delta}
T_{M}\Bigg(\frac{13+27w}{3+2w}\Bigg)-\frac{7w}{20\phi^2}q_{\beta\delta}\nabla_{\alpha}\phi\nabla^{\alpha}\phi+8\Big(\frac{\pi}{\phi}\Big)^{2}\Bigg[
TT_{\beta\delta}\Bigg(\frac{w+3}{3+2w}\Bigg)\right.\nonumber\\&-&\left.T^{2}q_{\beta\delta}\frac{(w^{2}+3w+3)}{(3+2w)^{2}}-2T^{\gamma}_{\;\delta}T_{\beta\gamma}+
q_{\beta\delta}T^{\alpha\gamma}T_{\alpha\gamma}\Bigg]-E_{\beta\delta}\right.\label{31}.
\ea Note that the value of $E_{\mu\nu}$ and the Brans-Dicke field,
as well as their derivatives, is not taken exactly on the brane,
but in the limiting $y\rightarrow 0^{\pm}$. Now we split the
matter energy-momentum in the form \cite{JP}
\be{T_{M\mu\nu}=-\Lambda
g_{\mu\nu}+\delta(y)T_{\mu\nu}},\label{32}\ee and
\be{T_{\mu\nu}=-\lambda q_{\mu\nu}+\tau_{\mu\nu}},\label{33}\ee
where $\Lambda$ is the cosmological constant of the bulk and
$\lambda$ the tension of the brane. It should be stressed that
such type of decomposition can lead to some ambiguity in the
cosmological scenario. However, here it can be done and, actually,
it is quite useful to interpret the final result. So, placing the
equations (\ref{32}) and (\ref{33}) inside the equation (\ref{31})
we obtain \ba
^{(5)}\!G_{\beta\delta}&=&\left.\frac{1}{2}\Bigg[\frac{1}{\phi}\nabla_{\nu}\nabla_{\sigma}\phi+\frac{w}{\phi^{2}}\nabla_{\nu}\phi\nabla_{\sigma}\phi
\Bigg](q_{\beta}^{\;\nu}q_{\delta}^{\;\sigma}-q^{\nu\sigma}q_{\beta\delta})+8\pi
\Omega\tau_{\beta\delta}-\Lambda_{5}q_{\beta\delta}\right.\nonumber\\&+&\left.8\Big(\frac{\pi}{\phi}\Big)^{2}\Sigma_{\beta\delta}-E_{\beta\delta}
\right. \label{34},\ea where
\be{\Omega=\frac{3\pi(w-1)\lambda}{\phi^{2}(3+2w)}}\label{35},\ee
\be{\Lambda_{5}=\frac{-4\pi\Lambda(21-41w)}{5\phi(3+2w)}+\Big(\frac{\pi}{\phi}\Big)^{2}\Bigg[\frac{7w}{20\pi^{2}}\nabla_{\alpha}\phi\nabla^{\alpha}\phi
+\frac{24(w-1)\lambda}{(3+2w)^{2}}[(w-1)\lambda+\tau]\Bigg]}\label{36}\ee
and
\be{\Sigma_{\beta\delta}=q_{\beta\delta}\tau^{\alpha\gamma}\tau_{\alpha\gamma}-2\tau^{\gamma}_{\;\delta}\tau_{\gamma\beta}+\Big(\frac{3+w}{3+2w}
\Big)\tau\tau_{\beta\delta}-
\frac{(w^{2}+3w+3)}{(3+2w)^{2}}q_{\beta\delta}\tau^{2}}.\label{37}\ee

Let's analyze the equation (\ref{34}) in more detail. First of
all, we do not write it in a Brans-Dicke form by the simple fact
that Brans-Dicke theory in not recovered on the brane in models
where the scalar field depends only on the extra dimension
\cite{AG}. We shall restrict the analysis to such cases. So, it
seems more plausible to look for deviations of the usual Einstein
brane-worlds formulation. Something like ``effective" Einstein's
equations on the brane. The first term arises from the scalar
field contribution and it brings information about the bulk
structure. Note that from the point of view of a brane observer,
such a field has no dynamics, since the brane is localized at a
fixed $y$. In the second term, there is a type of effective
Newton's coupling constant, which depends on the scalar field, as
well as on the brane tension. Since $\Omega=\Omega(\phi(r))$ the
scalar field must be stabilized in order to guarantee the
agreement with usual gravity on the brane. Artificially it can be
understood as an adjustment of the brane along to the extra
dimension, inducing the right value to the scalar field. Strictly
speaking, it should be done by the introduction of a well-behaved
potential in the Brans-Dicke part of the action, see for instance
ref. \cite{Perivo}.

As in \cite{JP}, here we recover the fact that it is not possible
to define a gravitational constant in an era where there was no
distinction between vacuum energy and usual matter energy.
Besides, the signal of $\Omega$ strongly depends on the signal of
the brane tension. The equation (\ref{36}) calls our attention
about the effective cosmological constant in five dimensions. It
depends on the bulk cosmological constant and on the scalar field,
hence it stresses the fact that the cosmological constant can be
variable for a bulk observer. The penultimate factor is quadratic
in the energy-momentum on the brane and could, hence, be an
important part of cosmological evolution in the early universe
(see, for example, \cite{COSMO} for the five dimensional case).
The last term provides more information about the gravitational
field of the bulk, which justifies the inclusion of the Weyl's
tensor in the analysis.

Returning to the first term of (\ref{34}), inserting the equations
(\ref{29}) and (\ref{30}) into (\ref{NC}) and taking into account
the split defined by equations (\ref{32}) and (\ref{33}), the
Codazzi equation gives an important relation between the scalar
field and derivatives of $\tau_{\mu\nu}$
\be{\frac{4\pi}{\phi}\Bigg[-D_{\nu}\tau_{\mu}^{\;\nu}+\frac{1}{3+2w}D_{\mu}\tau
\Bigg]=\Bigg[\frac{1}{\phi}\nabla_{\rho}\nabla_{\sigma}\phi+\frac{w}{\phi^{2}}\nabla_{\rho}\phi\nabla_{\sigma}\phi
\Bigg]n^{\sigma}q_{\mu}^{\;\rho}}.\label{38}\ee

As a last remark, we emphasize that the term $E_{\mu\nu}$, which
encodes the Weyl bulk tensor, also has restricted divergence by
derivatives of $\tau_{\mu\nu}$, as we can see by the contracted
Bianchi identities $D^{\beta}\;^{(5)}G_{\beta\delta}=0$ applied to
equation (\ref{34}),
\be{D^{\beta}E_{\beta\delta}=\frac{8\pi}{\phi}D^{\beta}\tau_{\beta\delta}-\frac{24\pi(w-1)\lambda}{(3+2w)^{2}}D_{\delta}\tau+
8\Big(\frac{\pi}{\phi}\Big)^{2}D^{\beta}\Sigma_{\beta\delta}}.\label{39}\ee
Obviously the two last equations can be used to relate the
Brans-Dicke field with the Weyl tensor. However it could not be
desirable if one wants to work with the extensive technology
developed for the $E_{\mu\nu}$ tensor (see Appendix A of
\cite{JP}).

\section{Conclusion}

In this work we generalized the Gauss-Codazzi formalism for
brane-worlds in Brans-Dicke gravity framework. It was done in the
context of $\mathbb{Z}_{2}$ symmetry models that allows us to
uniquely determine the extrinsic curvature on the brane. The main
result obtained is the equation (\ref{34}) which resembles the
results obtained for Einstein's theory, but it brings important
differences. The Brans-Dicke scalar field is present in all terms
of the right-hand side except in $E_{\mu\nu}$. Keeping in mind the
first term of (\ref{34}) together with equations (\ref{38}) and
(\ref{39}) it seems to be almost certain that we will find
shunting lines in the study the gravitational physical systems on
the brane as, for example, black hole area and quasar luminosity
\cite{ROY}. We shall address to those questions in the future.

Two interesting points call our attention in the equation
(\ref{34}). First, we remark the coupling between the brane
tension $\lambda$ and the Brans-Dicke parameter $w$. In the
standard formulation of the Brans-Dicke theory such parameter can
be expected to be of order of unity\footnote{Meanwhile experiments
shows that it is not the case. See for instance \cite{exp} for
current lower bound of the Brans-Dicke parameter.}. By the
equations (\ref{35}) and (\ref{36}) it is clear that in the case
of $(w=1)$ the effective Newton's constant vanishes, which is a
negative result, and there is no contribution of the brane tension
in the equation (\ref{34}). This type of inconsistence persists in
the non $\mathbb{Z}_{2}$ symmetric case \cite{vira}. However, we
should stress, it is a possible inconsistence between pure
Brans-Dicke theory and braneworld models, which is not the case
here, since we are using such theory in order to mimic low energy
gravity recovered from string theory. The second point is that
there is a configuration of the scalar field in which the induced
cosmological constant on the brane vanishes. From equation
(\ref{36}), assuming that the bulk cosmological constant is
constant, it is easy to see that $\Lambda_{5}=0$ if the scalar
field has the form
\ba\phi(y)=\frac{4BC+\Lambda^{2}A^{2}(y-D)^{2}}{4AB\Lambda},\label{nova}\ea
where $D$ is a constant of integration and
$A=\frac{4\pi(21-41w)}{5(3+2w)}$, $B=\frac{7w}{20}$,
$C=\frac{24\pi^2(w-1)\lambda[(w-1)\lambda+\tau]}{(3+2w)^2}$. We
note that $\Lambda_{5}$ also vanishes for
$\phi(y)=\frac{C}{A\Lambda}$. For nonconstant $\Lambda$, it is
necessary to have the explicitly behavior of the function in order
to do a similar analysis. Note that the constant solution for the
scalar field is not of physical interest for this type of
extension. The polynomial solution (\ref{nova}) is not usual in
the sense of models like \cite{AG}. However it is an important
result, since it can say, in the scope of the models in question,
what type of behavior of the Brans-Dicke scalar field can lead to
an effective cosmological constant on the brane.

It is important to remark the results obtained in the Appendix. It
is a direct generalization of the matching conditions to the
Brans-Dicke case. To conclude, we discuss a little bit more the
role played by the $\mathbb{Z}_{2}$ symmetry. The model we
consider has one compact on brane dimension. It is also
interesting to analyze this type of models without the
$\mathbb{Z}_{2}$ symmetry \cite{JPII}, since without such
simplification the final equations show a new term, that arise of
the mean of extrinsic curvature, which leads to an anisotropic
matter on the brane and then could be useful to interpret hybrid
brane world compactifications scenario.

\section*{Acknowledgments}

The authors benefited from useful discussions with Profs. Andrey
Bytsenko, Jos\'e Helay\"el-Neto and Hiroshi de Sandes Kimura. The
authors are also grateful to the EJPC referee for useful comments
and  enlightening viewpoints. J. M. Hoff da Silva thanks
CAPES-Brazil for financial support. M. E. X. Guimar\~aes and M. C.
B. Abdalla acknowledge CNPq for support.

\section*{Appendix: Israel-Darmois matching conditions to Brans-Dicke gravity}

In order to extend the usual junction conditions \cite{ID} to the
case in question we use distributional calculus, just like in
\cite{MAC}. The basic approach is to treat the brane as a
(infinitely thin) hypersurface orthogonally riddled by geodesics.
Denoting the extra dimensional coordinate by $y$, it is always
possible to choose some parametrization where the brane is located
at $y=0$, in such way that $y>0$ represents one side of the brane
and $y<0$ the other side. In this Appendix we use the notation
\be{[\chi]=\chi^{+}-\chi^{-}},\label{ap1}\ee where $\chi^{\pm}$
denote the limit of the $\chi$ quantity approaching the brane when
$y\rightarrow 0^{\pm}$. So, it is possible to analyze the jump of
$\chi$ through the hypersurface. With the Heaviside distribution
$\Theta(y)$ it is possible to decompose quantities at both sides
of the brane. The bulk metric can, then, be expressed as
\be{g_{\mu\nu}=\Theta(y)g_{\mu\nu}^{+}+\Theta(-y)g_{\mu\nu}^{-}
},\label{ap2}\ee where $g_{\mu\nu}^{+}$ ($g_{\mu\nu}^{-}$) is the
metric on the right (left) hand side. Note that the Heaviside
distribution has the following properties: $\Theta(y)=+1$ if
$y>0$, indeterminate if $y=0$ and zero otherwise. Besides \be{
\Theta^{2}(y)=\Theta(y)}, \nonumber \ee \be{
\Theta(y)\Theta(-y)=0},\ee \be{
\frac{d\Theta(y)}{dy}=\delta(y)}\label{ap3},\nonumber\ee where
$\delta(y)$ is the Dirac distribution. With such tools it is easy
to note that
\be{g_{\mu\nu,\;\alpha}=\Theta(y)g_{\mu\nu,\;\alpha}^{+}+\Theta(-y)g_{\mu\nu,\;\alpha}^{-}+\delta(y)[g_{\mu\nu}]n_{\alpha}
},\label{ap4}\ee and since the Christoffel symbols constructed
with (\ref{ap4}) will generate producs as $\Theta(y)\delta(y)$,
which is not well defined as a distribution, one is forced to
conclude that $[g_{\mu\nu}]=0$. This is the so-called Darmois
condition. In this Appendix we use the compact notation
$A_{,\;\mu}=\nabla_{\mu}A$, being $A$ any tensorial quantity.

Following this reasoning and supposing that the discontinuity of
$g_{\mu\nu,\;\alpha}$ is directed along $n^{\alpha}$ by
$[g_{\mu\nu,\;\alpha}]=k_{\mu\nu}n_{\alpha}$, for some tensor
$k_{\mu\nu}$, one finds that the left hand side (the geometrical
part) of the Einstein's tensor can be decomposed in a such way
that its $\delta$-function part (the part on the brane itself) is
given by \cite{MAC}
\be{\frac{1}{2}(k_{\gamma\mu}n^{\gamma}n_{\nu}+k_{\gamma\nu}n^{\gamma}n_{\mu}-kn_{\mu}n_{\nu}-k_{\mu\nu}-
(k_{\gamma\sigma}n^{\gamma}n^{\sigma}-k)g_{\mu\nu} )\equiv
S_{\mu\nu}}.\label{ap5}\ee

Let's now look at the Brans-Dicke part of the decomposition.
Writing the scalar field as
\be{\phi=\Theta(y)\phi^{+}+\Theta(-y)\phi^{-}}\label{ap6},\ee we
have
\be{\phi_{,\;\mu}=\Theta(y)\phi_{,\;\mu}^{+}+\Theta(-y)\phi_{,\;\mu}^{-}+\delta(y)[\phi]n_{\mu}}\label{ap7}.\ee
Since the Brans-Dicke equation is given by (\ref{6}) one has to
impose the continuity of the field across the brane, i. e.,
$[\phi]=0$, in order to avoid $\Theta(y)\delta(y)$ terms arising
from the second term of the hight hand side of (\ref{6}), just as
we did before. This is the analogous of the Darmois junction
condition to the Brans-Dicke case.

To find another matching condition on the brane, which involves
the extrinsic curvature, we need to decompose all terms of the
hight hand side of (\ref{6}) and equalize it to $S_{\mu\nu}$
(equation (\ref{ap5})). From (\ref{ap7}) we have
\be{\phi_{,\;\mu;\nu}=\Theta(y)\phi_{,\;\mu;\nu}^{+}+\Theta(-y)\phi_{,\;\mu;\nu}^{-}+\delta(y)[\phi_{,\;\mu}]n_{\nu}},\label{ap8}\ee
while the energy-momentum tensor is decomposed as
\be{T^{total}_{\mu\nu}=\Theta(y)T_{\mu\nu}^{+}+\Theta(-y)T_{\mu\nu}^{-}+\delta(y)T_{\mu\nu}},\label{ap9}\ee
where $T_{\mu\nu}$ is the energy-momentum on the brane. Before
substituting this factors into (\ref{6}) we have to deal with the
scalar fields in the denominator. Note that in the standard
distributional decomposition it is not difficult to see that
\be{\frac{1}{\phi}=\frac{\Theta(y)}{\phi^{+}}+\frac{\Theta{(-y)}}{\phi^{-}}}.\label{ap11}\ee
The $1/\phi^{2}$ factor obeys a similar splitting. So, after
substituting all decompositions into (\ref{6}) one finds that the
$\delta$-function part of the Brans-Dicke term is
\be{\frac{8\pi}{\phi}\Bigg(T_{\mu\nu}-\frac{g_{\mu\nu}}{3+2w}T\Bigg)
+\frac{1}{\phi}[\phi_{,\;\mu}]n_{\nu}\equiv
(BD)_{\mu\nu}}.\label{ap12}\ee Obviously
$S_{\mu\nu}=(BD)_{\mu\nu}$, and since $S_{\mu\nu}n^{\nu}=0$ (from
equation (\ref{ap5})) we arrive at
\be{[\phi_{,\;\mu}]=\frac{8\pi}{3+2w}T n_{\mu}}\label{ap13},\ee
since the contraction $T_{\mu\nu}n^{\nu}$ is zero (because
$T_{\mu\nu}$ belongs to the brane). The equation (\ref{ap13}) is
the analogous of $[g_{\mu\nu,\alpha}]=k_{\mu\nu}n_{\alpha}$. Now
substituting this result into $(BD)_{\mu\nu}$ we determine
completely the $\delta$-function term of the Brans-Dicke
equations, which is
\be{(BD)_{\mu\nu}=\frac{8\pi}{\phi}\Bigg(T_{\mu\nu}-q_{\mu\nu}\frac{T}{3+2w}\Bigg)
}.\label{ap14}\ee

The $S_{\mu\nu}$ term of Einstein's equation can be related with
the jump of the extrinsic curvature across the hypersurface
\cite{MAC} by $-([K_{\mu\nu}]-[K]q_{\mu\nu})$, then, after all we
have the analogous to the second matching condition for the
Brans-Dicke case given by
\be{\frac{8\pi}{\phi}\Bigg(T_{\mu\nu}-q_{\mu\nu}\frac{T}{3+2w}\Bigg)=-([K_{\mu\nu}]-[K]q_{\mu\nu})}.\label{ap15}\ee
This expression was obtained within the brane-world context.
However it is still valid for the usual four dimensions, if the
scalar field depends only on the transverse direction to the
hypersurface. We stress that, as expected, if $\phi=1/G$ and
$w\rightarrow \infty$ the expression for the Israel matching
condition in General Relativity is recovered.

\end{document}